\documentclass[twocolumn,amssymb,amsmath,aps,prl,groupedaddress,nobibnotes,showpacs]{revtex4}
\usepackage{graphicx}
\begin{document}

\title{Reliability of the beamsplitter based Bell-state measurement}

\author{Yoon-Ho Kim}\email{kimy@ornl.gov}
\affiliation{Center for Engineering Science Advanced Research, Computer Science \& Mathematics Division\\ Oak Ridge National Laboratory, Oak Ridge, Tennessee 37831}

\author{Warren P. Grice}
\affiliation{Center for Engineering Science Advanced Research, Computer Science \& Mathematics Division\\ Oak Ridge National Laboratory, Oak Ridge, Tennessee 37831}

\date{March 2003}

\begin{abstract}
A linear 50/50 beamsplitter, together with a coincidence measurement, has been widely used in quantum optical experiments, such as teleportation, dense coding, etc., for interferometrically distinguishing, measuring, or projecting onto one of the four two-photon polarization Bell-states $|\psi^{(-)}\rangle$. In this paper, we demonstrate that the coincidence measurement at the output of a beamsplitter cannot be used as an absolute identifier of the input state $|\psi^{(-)}\rangle$ nor as an indication that the input photons have projected to the $|\psi^{(-)}\rangle$ state.
\end{abstract}


\maketitle 

The phenomenon of non-local correlations, or entanglement, between quantum mechanical particles is central to the growing field of quantum information science. Entangled states have been used for experimentally verifying various violations of Bell's inequalities \cite{bell1,bell11,bell2}, as well as for demonstrations of quantum cryptography \cite{crypt}, quantum teleportation \cite{bouw,kim}, and quantum dense coding \cite{mattle}. In addition, the field of quantum computing relies on the ability to generate and manipulate multi-particle entangled states \cite{nielson}. Perhaps the simplest examples of entangled states are the polarization-entangled Bell states:
\begin{eqnarray}
|\psi^{(\pm)}\rangle &=& (|H\rangle_1|V\rangle_2 \pm |V\rangle_1|H\rangle_2)/\sqrt{2},\nonumber\\
|\phi^{(\pm)}\rangle &=& (|H\rangle_1|H\rangle_2 \pm |V\rangle_1|V\rangle_2)/\sqrt{2},\nonumber
\end{eqnarray}
where $|H\rangle$ and $|V\rangle$ refer to the horizontal and vertical polarization states of a single-photon, respectively. Such states are routinely generated via the process of spontaneous parametric down-conversion (SPDC) \cite{bell2,bell3,bell4}.

Since the Bell-states form a complete (entangled) basis for the two-particle polarization Hilbert space, it should be possible to build a measurement device capable of distinguishing all four Bell states. Although the Bell-state measurement (BSM) plays a critical role in many of the quantum applications mentioned above, it is not trivial to build such a device, as non-linear photon-photon interactions are required for a complete BSM \cite{bsm2}. Thus far, there has been only one experimental demonstration of a complete BSM (for teleportation) using non-linear optical effects \cite{kim}. On the other hand, a simple linear optical beamsplitter has been claimed to distinguish at least one out of four Bell-states \cite{bsm} and has been used in several recent experiments \cite{mattle,bouw}.

\begin{figure}[b]
\includegraphics[width=3.in]{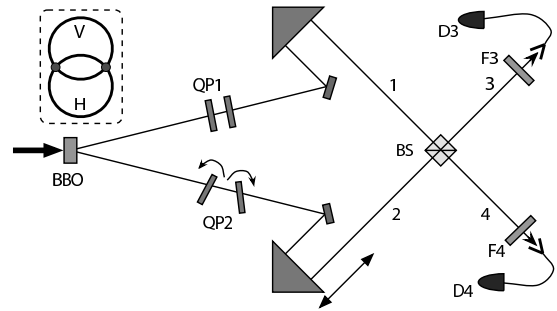}
\caption{\label{fig:setup}Outline of Experimental setup. A 3 mm thick type-II BBO crystal is pumped by a 120 fsec ultrafast laser pulse.}
\end{figure}

The beamsplitter based BSM can be briefly explained as follows. Consider a 50/50 beamsplitter in which two photons in a Bell state enter via modes 1 and 2 and exit via modes 3 and 4 (see the beamsplitter BS in Fig.~\ref{fig:setup}). It is straightforward to show that, out of the four Bell-states, only the $|\psi^{(-)}\rangle_{1,2}$ input results in exactly one photon in each output port \cite{bsm}. Assuming perfect detectors, therefore, the probability of a coincidence count between two detectors located at modes 3 and 4 is unity for the $|\psi^{(-)}\rangle_{1,2}$ input state. For the other three Bell-states, the probability of coincidence is zero because both photons always end up either in mode 3 or in mode 4. 

Experimentally, the presence of a coincidence or null-coincidence can be confirmed by varying the overlap of the `photon wavepackets' at the beamsplitter. If photons do not overlap at the beamsplitter, they scatter randomly and the probability of (background) coincidence is 1/2. When the paths are properly aligned, the state $|\psi^{(-)}\rangle_{1,2}$ produces a peak in the coincidence rate that is twice the background coincidence rate. Likewise, the other three Bell-states produce a dip in the coincidence rate as the photon overlap at the beamsplitter is varied. The presence of these coincidence features is often regarded as evidence that a particular apparatus is properly aligned and is functioning as a BSM device. In this paper, we show that these features may be observed even when Bell states are not used as the inputs. Thus, the presence of a coincidence peak does not guarantee that the input state is $|\psi^{(-)}\rangle_{1,2}$. The implication, therefore, is that a coincidence event cannot be used as an absolute indication that an unknown input state has collapsed to $|\psi^{(-)}\rangle_{1,2}$.

The experimental setup is shown in Fig.~\ref{fig:setup}. A 3 mm thick type-II BBO crystal is pumped by a train of 120 fsec ultrafast pulses centered at 390 nm. Photons centered at 780 nm are emitted into two distinct cones, one corresponding to the e-ray (V-polarized) and the other to the o-ray (H-polarized) of the crystal. Interest is restricted to the intersections of the two light cones, shown in the inset, where photons of either polarization may be found. Before being directed to the input ports of an ordinary non-polarizing beamsplitter, the photons pass through 600$\mu$m thick quartz plates QP1 and QP2, which are oriented with their optic axes parallel to that of the BBO crystal. The quartz plates are used to adjust the phase between the interfering terms, as described below. After exiting the beamsplitter, the photons are detected by single-photon counters, D3 and D4, and the coincidence rate is measured using a time-to-amplitude converter and a multi-channnel analyzer with an effective coincidence window of 3 nsec. F3 and F4 are 20 nm (FWHM) spectral filters centered at 780 nm.

The two-photon state exiting the quartz plates may be written in simple form as
\begin{equation}
|\psi\rangle = (|H(t_H)\rangle_1|V(t_V)\rangle_2 + e^{-i \phi}|V(t_V)\rangle_1|H(t_H)\rangle_2)/\sqrt{2},\nonumber
\end{equation}
where, for example, $|H(t_H)\rangle_1|V(t_V)\rangle_2$ represents a horizontally polarized photon in path 1 and a vertically polarized photon in path 2 with the most probable times of emission being $t_H$ and $t_V$, respectively. The photon wave packets are centered at different times because they propagate through the birefringent materials at different speeds. In most experiments involving this type of source, birefringent plates are used to temporally overlap the orthogonally polarized photons \cite{bell3,rubin}. No such compensation is present here, however. The relative phase $\phi$ between the two terms is determined by the transit times for the orthogonally polarized photons in the two sets of quartz plates. Tilting the plates in one arm increases the effective thickness of the plates, permitting precise phase adjustment. 

\begin{figure}[t]
\includegraphics[width=3.in]{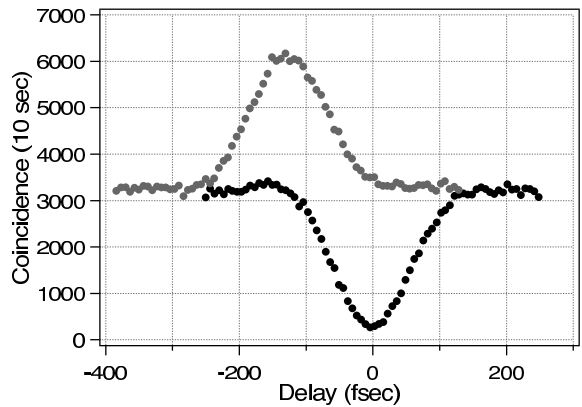}
\caption{\label{fig:data}Experimental data. The peak-dip visibility is $91$\%.}
\end{figure}

The coincidence data is shown in Fig.~\ref{fig:data}, with the two different data sets corresponding to two different phase settings (0 and $\pi$). The phase is adjusted by tilting QP2. This adjustment also increases the total effective path in the lower arm, an effect which is manifested as an offset between the peak and the dip. 

In spite of the fact that the data clearly shows the coincidence peak and dip typically associated with BSM, the input states are \textit{not} Bell states: polarization correlation measurements performed here would not yield the high-visibility sinusoidal curves associated with polarization-entangled states and, consequently, these polarization states could not be used to violate Bell's inequality. The lack of entanglement here is due not only to the problems typically associated with ultrafast-pumped type-II sources, see Ref.~\cite{ultrafast1,kimgrice1,ultrafast2}, but also to the fact that the orthogonally polarized photons are not temporally overlapped, as described above. 

\begin{figure}[t]
\includegraphics[width=3.1in]{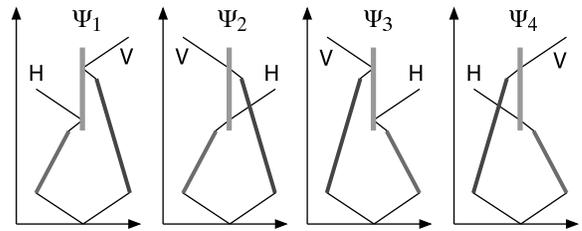}
\caption{\label{fig:amplitudes}Four Feynman alternatives occur in this experiment. Vertical gray line represents the beamsplitter. For simplicity, the net delays (occured in the SPDC crysal and in the quartz) are expressed as  thick lines in the two-photon paths. }
\end{figure}

The data shown in Fig.~\ref{fig:data} shows, therefore, that polarization entanglement is not required at the beamsplitter input to observe the coincidence peak (or dip) at the beamsplitter output. The effect, which is discussed in more detail elsewhere \cite{kimgrice4}, can be understood in terms of the Feynman diagrams for the events leading to a coincidence detection. There are two ways (corresponding to the two terms in the two-photon state) that photons may be emitted into the two input arms. A coincidence may be recorded either when both photons are transmitted (t-t) or when both are reflected (r-r), for a total of four Feynman amplitudes, as shown in Fig.~\ref{fig:amplitudes}. For a given emission event ($|H(t_H)\rangle_1|V(t_V)\rangle_2$, for example), the r-r and t-t cases are distinguishable, since they lead to different sequences of detection events (compare $\Psi_1$ and $\Psi_2$ in Fig.~\ref{fig:amplitudes}). As long as the two arms of the interferometer are identical, though, a particular detection sequence may be obtained via two distinct emission events (compare $\Psi_1$ and $\Psi_4$ in Fig.~\ref{fig:amplitudes}), i.e., the amplitudes are pairwise indistinguishable. Depending on the phase between the two emission terms, the resulting interference may be either constructive or destructive.

Interference curves similar to those shown in Fig.~\ref{fig:data} are typically used to align a BSM device. It is then assumed that an unknown input state is projected to $|\psi^{(-)}\rangle_{1,2}$ whenever a coincidence is observed at the outputs. We have shown here that the same curves may be obtained with states that are not Bell states. It follows, then, that a coincidence detection does not necessarily project the input state to $|\psi^{(-)}\rangle_{1,2}$. Rather, the coincidence measurement projects the input state to a class of states that possess a particular symmetry and, as we shall see below, the Bell state $|\psi^{(-)}\rangle_{1,2}$ is just one of many two-photon states which exhibit such symmetry.

The symmetry condition mentioned above can be identified by determining the input state that results exclusively in the two photons exiting the beamsplitter via different paths, i.e., the input state that always leads to a coincidence detection (coincidence peak). We start by considering the more general two-photon state
\begin{eqnarray}
|\psi\rangle &=& \iint d\omega_H \, d\omega_V \{ \mathcal{F}_{H1,V2}(\omega_H,\omega_V) \hat a_{H1}^\dagger(\omega_H) \hat a_{V2}^\dagger (\omega_V) \nonumber \\ &+& \mathcal{F}_{V1,H2}(\omega_H,\omega_V) \hat a_{V1}^\dagger(\omega_V) \hat a_{H2}^\dagger (\omega_H) \} |0\rangle/\sqrt{2}, \label{eq:input}
\end{eqnarray}
where, for example, $\hat a_{H1}^\dagger(\omega_H) \hat a_{V2}^\dagger (\omega_V)|0\rangle$ represents a horizontally polarized single photon of frequency $\omega_H$ in path 1 and a vertically polarized single photon of frequency $\omega_V$ in path 2. The two-photon joint spectral function $\mathcal{F}_{H1,V2}(\omega_1,\omega_2)$ describes the energy distribution probabilities for the photon pair and can be calculated explicitly in the case of SPDC \cite{rubin,ultrafast1,kimgrice1}. The most general two-photon state should also include terms of the form $\hat a_{H1}^\dagger(\omega_1) \hat a_{H2}^\dagger(\omega_2)$ and $\hat a_{V1}^\dagger(\omega_1) \hat a_{V2}^\dagger(\omega_2)$, but since these terms always lead to the symmetry condition in which two photons exit via the same output path (null coincidence or coincidence dip), we can restrict attention to the terms shown in Eq.~(\ref{eq:input}) without loss of generality. 

 The input and output modes of the beamsplitter are related by
$
\hat a_{j3}(\omega) = \left[ \hat a_{j2}(\omega) + i \hat a_{j1}(\omega) \right]/\sqrt{2}$ and
$
\hat a_{j4}(\omega) = \left[ \hat a_{j1}(\omega) + i \hat a_{j2}(\omega) \right]/\sqrt{2},$ where the subscript $j$ identifies the polarization (H or V). Inserting these operators into Eq.~(\ref{eq:input}) yields the state at the output of the beamsplitter,
\begin{widetext}
\begin{eqnarray}
|\psi\rangle_{3,4}=\frac{1}{2}\iint d\omega_H \, d\omega_V \left\{
[ \mathcal{F}_{H1,V2}(\omega_H,\omega_V) - \mathcal{F}_{V1,H2}(\omega_H,\omega_V) ] [ \hat a_{H4}^\dagger(\omega_H) \hat a_{V3}^\dagger(\omega_V) - \hat a_{H3}^\dagger(\omega_H) \hat a_{V4}^\dagger(\omega_V) ]\right. \nonumber\\  
\left. + i [ \mathcal{F}_{H1,V2} (\omega_H,\omega_V) + \mathcal{F}_{V1,H2}(\omega_H,\omega_V) ] [ \hat a_{H3}^\dagger(\omega_H) \hat a_{V3}^\dagger(\omega_V) + \hat a_{H4}^\dagger(\omega_H) \hat a_{V4}^\dagger(\omega_V) ] \right\} 
|0\rangle. \label{eq:output}
\end{eqnarray}
\end{widetext}
If the input state is to lead to exactly one photon in each of the output paths, then the coefficients preceding operators of the forms $\hat a_{H3}^\dagger(\omega_H) \hat a_{V3}^\dagger(\omega_V)$ and $\hat a_{H4}^\dagger(\omega_H) \hat a_{V4}^\dagger(\omega_V)$ must be zero. This leads to the condition $\mathcal{F}_{H1,V2}(\omega_1,\omega_2) = - \mathcal{F}_{V1,H2}(\omega_2,\omega_1)$. Imposing this condition on Eq.~(\ref{eq:input}) yields the anti-symmetric state
\begin{eqnarray}
|\psi\rangle_{AS} &=& \iint d\omega_H \, d\omega_V \mathcal{F}(\omega_H,\omega_V) \{ \hat a_{H1}^\dagger(\omega_H) \hat a_{V2}^\dagger (\omega_V) \nonumber \\ &-& \hat a_{V1}^\dagger(\omega_V) \hat a_{H2}^\dagger (\omega_H) \} |0\rangle/\sqrt{2}, \label{eq:antisymm}
\end{eqnarray}
which has the property that the photons' \textit{spectral and temporal properties are correlated with their polarizations}. This condition is satisfied in the type-II emission scheme employed here, as long as the optical path lengths in the two arms are identical (path length mismatch is manifested as a frequency dependent phase factor).

The symmetry exhibited in Eq.~(\ref{eq:antisymm}), while sufficient for deterministically generating a coincidence at the beamsplitter output, does not guarantee that the state is a Bell state. This can be seen by analyzing the polarization correlations of the two photons \cite{bell11,bell2,bell3,bell4}. If a pair of photons in a polarization-entangled state are directed to detectors preceded by polarizers, the coincidence rate will vary sinusoidally with either the sum or difference of the polarizer angles. Any state exhibiting this type of correlation may be used to violate a Bell inequality. For the $|\psi^{(-)}\rangle_{1,2}$ state, the coincidence rate is given by $R_c \propto \sin^2(\theta_1-\theta_2)$, where $\theta_1$ and $\theta_2$ are the orientations of the polarizers. The coincidence rate for the general state $|\psi\rangle$ is given by
\begin{equation}
R_c(\omega_1,\omega_2) \propto |\langle\theta_1(\omega_1)|\langle\theta_2(\omega_2)|\psi\rangle |^2,\nonumber
\end{equation}
where $\langle\theta_j(\omega_j)|=\langle0|(\cos\theta_j \hat a_{Hj}(\omega_j)+\sin\theta_j \hat a_{Vj}(\omega_j))$. With the state $|\psi\rangle$ given in Eq.~(\ref{eq:input}), the coincidence rate becomes
\begin{eqnarray}
R_c(\omega_1,\omega_2) &\propto& |\cos\theta_1\sin\theta_2 \,\mathcal{F}_{H1,V2}(\omega_1,\omega_2)\nonumber\\  &+& \cos\theta_2\sin\theta_1 \,\mathcal{F}_{V1,H2}(\omega_2,\omega_1)|^2,\label{eq:polcorr}
\end{eqnarray}
which is proportional to $\sin^2(\theta_1-\theta_2)$ only if $\mathcal{F}_{H1,V2}(\omega_1,\omega_2) = - \mathcal{F}_{V1,H2}(\omega_1,\omega_2)$. As before, this condition may be imposed on Eq.~(\ref{eq:input}) to give the Bell state
\begin{eqnarray}
|\psi\rangle_{Bell} &=& \iint d\omega_1 \, d\omega_2 \mathcal{F}(\omega_1,\omega_2) \{ \hat a_{H1}^\dagger(\omega_1) \hat a_{V2}^\dagger (\omega_2) \nonumber \\ &-& \hat a_{V1}^\dagger(\omega_1) \hat a_{H2}^\dagger (\omega_2) \} |0\rangle, \label{eq:entangled}
\end{eqnarray}
where the different labeling scheme reflects a symmetry that is subtly different than that shown in Eq.~(\ref{eq:antisymm}). Here, \textit{the spectral and temporal properties of the photons are correlated with path}, rather than with polarization. For the Bell-state, therefore, the horizontally polarized photon in a particular path must be identical to the vertically polarized photon in that path. This condition is not satisfied for the photon pair source employed here, not only because the orthogonally polarized photons are centered at different times \cite{rubin}, but also because the different spectral properties of the emitted photons are correlated with polarization \cite{ultrafast1,kimgrice1}. However, the symmetry condition shown in Eq.~(\ref{eq:entangled}) can be met if the two-photon state is ``rearranged'' so that any properties originally correlated with polarization become correlated, instead, with path \cite{bell4}.

These results are summarized as follows: a two-photon state with the symmetry of Eq.~(\ref{eq:antisymm}) will produce a coincidence at the beamsplitter output, while a state with the symmetry of Eq.~(\ref{eq:entangled}) will exhibit the polarization correlations of a Bell state. Of course it is possible for a state to possess both types of symmetry, in which case the beamsplitter really would identify the $|\psi^{(-)}\rangle$ Bell state . Both symmetry conditions are met if the photons' spectral and temporal properties are correlated with neither path nor polarization, i.e. if $\mathcal{F}(\omega,\omega') = \mathcal{F}_(\omega',\omega)$. In this case, the two photons are spectrally and temporally identical, as is the case with the photon pair source described in Ref.~\cite{bell3}. For ultrafast-pumped type-II SPDC used in this paper, this condition may be satisfied by  configuring it to eliminate spectral differences between the photons \cite{kimgrice1, griceuren}.

The analysis presented above shows that the symmetry condition that leads to a coincidence at the beamsplitter output is different than the symmetry condition required for polarization entanglement. Although the analysis was carried out in the spectral domain, equivalent results would be obtained in the time domain, where the improperly comensated temporal walk-off would be represented as a temporal shift, rather than an additional phase factor. It has been assumed here that the emitted photons, while not spectrally identical, have identical center wavelengths. Nothing in the analysis, however, requires this to be so and the differences in the symmetry conditions may best be illustrated by considering a two-color two-photon source. Imagine a source that emits one red and one blue photon into two distinct paths and that either polarization may be found in each path, with the polarizations always found to be orthogonal when measured in the H-V basis. Depending on whether photon color is correlated with polarization or with path, such a source may: i) have unit probability of producing a coincidence count (coincidence peak) at the beamsplitter output while exhibiting no polarization entanglement, e.g., $(|H_R\rangle_1|V_B\rangle_2 - |V_B\rangle_1|H_R\rangle_2)/\sqrt{2}$ (the same result as in the experiment presented here); or ii) be polarization-entangled ($|\psi^{(-)}\rangle_{1,2}$ state), but fail to produce a coincidence peak at the beamsplitter output, e.g., $(|H\rangle_{1R}|V\rangle_{2B} - |V\rangle_{1R}|H\rangle_{2B})/\sqrt{2}$. In the latter case, the red photon is always found to be in path 1, while the blue photon is always in path 2 \cite{last}. The pair would be entangled in polarization (assuming no additional timing information), but when incident on a beamsplitter the photons would not exhibit the interference features shown in Fig.~\ref{fig:data}, since the coincidence detection events would no longer be pairwise indistinguishable as in Fig.~\ref{fig:amplitudes}.

In conclusion, we have presented experimental evidence that a successful Bell-state measurement cannot be claimed solely based on the coincidence data alone, because the interference features (coincidence peak or dip) which are commonly considered as the signature of a successful BSM may, in fact, be obtained with input states incapable of violating a Bell inequality. We have also shown that the conditions which lead to a positive result (coincidence at the beamsplitter outputs) are indeed different than the conditions required of a polarization-entangled state or a proper Bell-state projection. 

The authors wish to acknowledge several enlightening discussions with D. Branning and helpful comments by V. Protopopescu. This research was supported in part by the National Security Agency and the LDRD Program of Oak Ridge National Laboratory, managed for the U.S. DOE by UT-Battelle, LLC, under contract No.~DE-AC05-00OR22725.


\begin{thebibliography}{}
\vspace*{-0.4cm}

\bibitem{bell1} J.S. Bell, Physics \textbf{1}, 195 (1964); J.F. Clauser and A. Shimony, Rep. Prog. Phys. \textbf{41}, 1881 (1978).

\bibitem{bell11} A. Aspect, P. Grangier, and G. Roger, Phys. Rev. Lett. \textbf{47}, 460 (1981).

\bibitem{bell2} Y.H. Shih and C.O. Alley, in \textit{Proc. 2nd Int. Symp. Found. Quantum Mechanics}, ed. M. Namiki (Physical Society of Japan, Tokyo, 1987); Phys. Rev. Lett. \textbf{61}, 2921 (1988); Z.Y. Ou and L. Mandel, Phys. Rev. Lett. \textbf{61}, 50 (1988). 

\bibitem{crypt} N. Gisin \textit{et al.}, Rev. Mod. Phys. \textbf{74}, 145 (2002).

\bibitem{bouw} D. Bouwmeestter \textit{et al.}, Nature \textbf{390}, 575 (1997); J.-W. Pan \textit{et al.}, Phys. Rev. Lett. \textbf{80}, 3891 (1998). 

\bibitem{kim} Y.-H. Kim, S.P. Kulik, and Y. Shih, Phys. Rev. Lett. \textbf{86}, 1370 (2001); J. Mod. Opt. \textbf{49}, 221 (2001).

\bibitem{mattle} K. Mattle \textit{et al.}, Phys. Rev. Lett. \textbf{76}, 4656 (1996)

\bibitem{nielson} M.A. Nielson and I.L. Chuang, \textit{Quantum Computation and Quantum Information} (Cambridge Univ. Press 2000).

\bibitem{bell3} P.G. Kwiat \textit{et al.} Phys. Rev. Lett. \textbf{75}, 4337 (1995). 

\bibitem{bell4} Y.-H. Kim \textit{et al.} Phys. Rev. A \textbf{67}, 010301(R) (2003).

\bibitem{bsm2} L. Vaidman and N. Yoran, Phys. Rev. A \textbf{59}, 116 (1999); N. L\"{u}tkenhaus, J. Calsamiglia, and K.-A. Suominen, \textit{ibid.}, \textbf{59}, 3295 (1999). 

\bibitem{bsm} S.L. Braunstein and A. Mann, Phys. Rev. A \textbf{51} R1727 (1995).

\bibitem{rubin} M.H. Rubin, D.N. Klyshko, Y.H. Shih, and A.V. Sergienko, Phys. Rev. A
\textbf{50}, 5122 (1994).

\bibitem{ultrafast1} T.E. Keller and M.H. Rubin, Phys. Rev. A \textbf{56}, 1534 (1997); W.P. Grice and I.A. Walmsley, Phys. Rev. A \textbf{56}, 1627 (1997).

\bibitem{kimgrice1} Y.-H. Kim and W.P. Grice, J. Mod. Opt. \textbf{49}, 2309 (2002).

\bibitem{ultrafast2} G. Di Giuseppe \textit{et al.}, Phys. Rev. A 56, R21(1997); W.P. Grice \textit{et al.}, Phys. Rev. A \textbf{57}, R2289 (1998); Y.-H. Kim \textit{et al.}, Phys. Rev. A \textbf{64}, 011801(R) (2001); Y.-H. Kim \textit{et al.}, Phys. Rev. Lett. \textbf{86}, 4710 (2001).

\bibitem{kimgrice4} Y.-H. Kim and W.P. Grice, to be published.

\bibitem{griceuren} W.P. Grice, A.B. U'Ren, and I.A. Walmsley, Phys. Rev. A \textbf{64}, 063815 (2001).

\bibitem{last} Y.-H. Kim, S.P. Kulik, and Y. Shih, Phys. Rev. A \textbf{63}, 060301(R) (2001).

\end{thebibliography}
\end{document}